\documentclass[10pt,conference]{IEEEtran}

\usepackage{amsmath,amssymb}
\usepackage{graphicx,cite}
\usepackage{epstopdf}
\usepackage{url}
\usepackage{color}

\begin{document}
\title{Classifying flows and buffer state for \\YouTube's HTTP adaptive streaming service in mobile networks}

\author{\IEEEauthorblockN{Dimitrios Tsilimantos, Theodoros Karagkioules, and Stefan Valentin}
	\IEEEauthorblockA{
		Mathematical and Algorithmic Sciences Lab, Paris Research Center\\  
		Huawei Technologies France\\ 
		\{dimitrios.tsilimantos, theodoros.karagkioules, stefan.valentin\}@huawei.com}}

\maketitle

\begin{abstract}
Accurate cross-layer information is very useful to optimize mobile networks for specific applications. However, providing application-layer information to lower protocol layers has become very difficult due to the wide adoption of end-to-end encryption and due to the absence of cross-layer signaling standards. As an alternative, this paper presents a \emph{traffic profiling} solution to passively estimate parameters of HTTP Adaptive Streaming (HAS) applications at the lower layers. By observing IP packet arrivals, our machine learning system identifies video flows and detects the state of an HAS client's play-back buffer in real time. Our experiments with YouTube's mobile client show that Random Forests achieve very high accuracy even with a strong variation of link quality. Since this high performance is achieved at IP level with a small, generic feature set, our approach requires no Deep Packet Inspection (DPI), comes at low complexity, and does not interfere with end-to-end encryption. Traffic profiling is, thus, a powerful new tool for monitoring and managing even encrypted HAS traffic in mobile networks.
\end{abstract}

\begin{IEEEkeywords}
	HTTP Adaptive Streaming; YouTube; MPEG-DASH; Service Classification; Machine Learning.
\end{IEEEkeywords}

\section{Introduction}
\bstctlcite{IEEEbibChanges:BSTcontrol}
\label{sec:intro}
Mobile video streaming generates a significant portion of traffic in cellular networks. According to Cisco's traffic report  \cite{cisco2016}, video already represented 60\% of the mobile IP traffic in 2016 and is projected to reach 78\% by 2021. This traffic is dominated by HTTP Adaptive Streaming (HAS) services \cite{Sandvine}, which follow the Dynamic Adaptive Streaming over HTTP (DASH) standard \cite{DASH} or the HTTP Live Streaming (HLS) specification \cite{HLS}.

Reacting to this heavy increase in video traffic, Mobile Network Operators (MNOs) have started to deploy traffic shaping solutions. In November 2015, T-Mobile USA deployed \emph{BingeOn} which offers an unlimited plan for video streaming while throttling the video bit-rate to ``approximately 1.5 Mbit/s averaged over one minute of video'' \cite{BingeOnTechSpec}. A similar solution became operational in Germany in April 2017 under the term \emph{StreamOn}. Other MNOs investigate similar solutions for traffic shaping, while network equipment vendors are customizing base station schedulers to support video streaming by specific rate guarantees \cite{ourICC16} or weight adjustment \cite{AGBRpaper}.

All these solutions require a certain degree of application-layer knowledge. Before a video flow can be throttled or scheduled with specific rules, its packets need to be identified. Once identified, it is beneficial to know the video bit-rate and play-back buffer state of that stream in real time. This information allows schedulers, traffic shaping and admission control schemes to minimize their impact on Quality of Experience (QoE) or to even increase it by providing bit-rate guarantees when possible \cite{AGBRpaper}.

This demand for accurate application-layer information is a major practical problem. Network optimization functions typically operate at the Layer 2 and 3 of the ISO/OSI protocol stack, while application information is available at Layer 7. Currently, MNOs solve this cross-layer signaling problem by a combination of explicit signaling or Deep Packet Inspection (DPI). DPI aims to dissect Layer 7 traffic at Layer 2 or 3 and extracts flags and parameters from protocol headers above Layer 3 or even from payload. Explicit signaling, however, requires Over-The-Top (OTT) media services to add specific flags or tags to their streaming data. 

Since these methods are based on simple rule systems and string comparison, they are susceptible to spoofing. A malicious user can mislead a flow classifier by injecting the appropriate flags or tags in its own traffic, which may not necessarily be a video stream. By using an HTTP proxy, a simple spoofing method for BingeOn was demonstrated in \cite{bingeonPaper} only several months after this service became available. In addition to spoofing, the use of end-to-end encryption techniques such as Transport Layer Security (TLS) and Secure Sockets Layer (SSL) is a major issue for DPI. Since most major streaming OTTs have adopted TLS/SSL, MNOs cannot directly apply DPI to such encrypted video flows. In order to access the application-layer information, they have to interrupt the end-to-end encryption, e.g., by terminating the end-point on the client side. Since the OTT has to accept such ``man in the middle attacks'', this approach weakens operational security, leads to rejected certificates with more rigorous clients, and eventually opens the door for malicious use.

This paper tackles the cross-layer signaling problem only by observing the IP packet flows of HAS traffic. We follow the main idea of \emph{traffic profiling} by estimating application-layer information based on characteristics observed at the lower layers. In particular, we observe information such as the IP addresses of source and destination, IP packet size, and IP packet arrival time from queues in the network. Based on this input, our system separates HAS video flows from non-HAS traffic and estimates the current state of the video client's play-back buffer. Based on 120 hours of end-to-end encrypted traffic data from YouTube, our approach performs this classification at very high accuracy. This is a consequence of the regular arrival patterns of HAS, our careful feature design and the use of state-of-the-art machine learning methods.

The particular contributions of our paper are:
\begin{enumerate}
	\item A new traffic profiling system that classifies the flow type and play-back buffer state of HAS at the IP layer in real time.
	\item A careful feature design that is generic enough to equally work with TCP and UDP-based streaming and leads to a small feature set for low computational complexity.
	\item A rigorous experimental design to collect data and ground truth from YouTube's video streaming service. 
	\item A representative performance evaluation with interesting insights on the effect of the selected features, machine learning method and link quality.
\end{enumerate} 

The remainder of this paper is organized as follows. We discuss related work in Section \ref{sec:relatedWork} and summarize our system assumptions in Section \ref{sec:systemModel}. Then, we present our traffic profiling solution in Section \ref{sec:TPmodel} including the machine learning system. We detail the methodology of our YouTube experiments in Section \ref{sec:ExpMethodology}, present experimental results in Section \ref{sec:results}, and conclude the paper in Section \ref{sec:conclusions}.
 
\section{Related work} 
\label{sec:relatedWork}
The main idea of using passive traffic measurements for recognizing statistical patterns of video traffic has been studied in several works, as for example in  \cite{surveyMLclassification,passiveDimopoulos,Orsolic2017,Dimopoulos,Katsarakis,AudioVideo,BUFFEST}. Besides \cite{passiveDimopoulos} that focuses on the analysis and understanding of the obtained measurement data, the rest of the studies rely on Machine Learning (ML)  methods, which have become prominent for traffic classification. We encourage interested readers to refer to \cite{surveyMLclassification} and the references therein for a detailed comprehensive survey. 

When it comes to video streaming, most of the recent works propose to distinguish the entire video session into different classes. For example, the authors in \cite{Orsolic2017} propose a system that monitors application-level quality indicators and corresponding traffic traces to classify YouTube videos into three QoE classes. Similarly, different levels of QoE are studied in \cite{Dimopoulos} with the focus on stalling, average video quality and quality variations as the key influence factors. A causal analysis between QoE and Quality of Service (QoS) is presented in \cite{Katsarakis} with features from application and network-layer QoS metrics, while an approach to discriminate between audio and video HAS flows is proposed in \cite{AudioVideo}. Compared to these studies that classify an entire video session into a single category, our classification is performed at a higher temporal resolution by estimating dynamic video traffic parameters in real time.

Closer to our work is the methodology presented in \cite{BUFFEST}, where the target is to predict the class of the play-back buffer level during the video session by defining a set of buffer levels in seconds. Unlike in this work, our focus is the prediction of the buffer state, which is a more fundamental property of adaptive streaming clients. 

Our own prior work points to an interesting application for buffer state classification. In \cite{ourTP_ICC17}, we exploited that every adaptive streaming client strives to achieve a rate match between the client's download rate (i.e., throughput) and the server's source rate (i.e., content encoding rate). This match is obtained in the steady state where, consequently, throughput is a good predictor for encoding rate. This rate estimation requires the accurate detection of the steady state, which we achieved by simple heuristics in \cite{ourTP_ICC17}. As this fixed-rule approach failed for the more complicated variation of link quality in mobile networks \cite{Forum2014}, we now adopt ML models to generalize buffer state detection to a wider set of practical streaming scenarios and to perform HAS traffic classification as well.

\section{System model} 
\label{sec:systemModel}
A packet flow is defined as a series of packets sharing the same source and destination IP addresses, source and destination IP ports and transport protocol. Without loss in generality, we assume that packets are generated with a typical TCP/IP stack, i.e., TCP or UDP is used on top of IP. Packet flows are also distinguished in different classes, each class indicating IP traffic that belongs to a specific application, e.g. Web, file transfer, gaming, Peer-to-Peer (P2P), chat and media streaming. During video streaming, the source node is a video server and the destination node is a video client software running on a mobile device. HAS is considered according to standards \cite{DASH,HLS} and the communication between server and client utilizes at least one intermediate node, e.g. edge-router, gateway, Base Station (BS) or access point, close to the edge that forwards the packets in either direction. This last assumption is necessary since our traffic profiling solution is based on the fact that the observed packet Inter-Arrival Times (IATs) are close to end-to-end IATs. Multiplexing in the same packet flow is allowed in our system, as long as it covers only different audio and video streams of a single HAS session and user. This enables us to include Google's Quick UDP Internet Connections (QUIC) \cite{quic} in our studies, a multiplexed stream transport protocol over UDP that is widely used lately. 

Briefly describing an HAS system, a play-back buffer is used at the client side in order to compensate for variations in received throughput, due to the dynamic nature of the wireless channel conditions, but also in video encoding rate, as Variable Bit-Rate (VBR) is commonly adopted for encoding. Moreover, a video is divided into a sequence of smaller segments containing a short interval of content play-back time, typically in the order of few seconds. Video segments are encoded in multiple quality representations, which are stored in the video server. The client is then able to adjust the play-out quality by sequentially requesting segments in the representation indicated by the algorithm of the deployed HAS policy, which usually takes into account buffer level information and throughput statistics. While the description of different HAS policies goes beyond the scope of this paper, recent insightful comparative studies can be found in \cite{TheoNOSSDAV17,HASpolicies}.

\begin{figure}[t]
	\centering
	\includegraphics[width=1\linewidth,trim={0.1cm 0.2cm 0.2cm 0.2cm},clip]{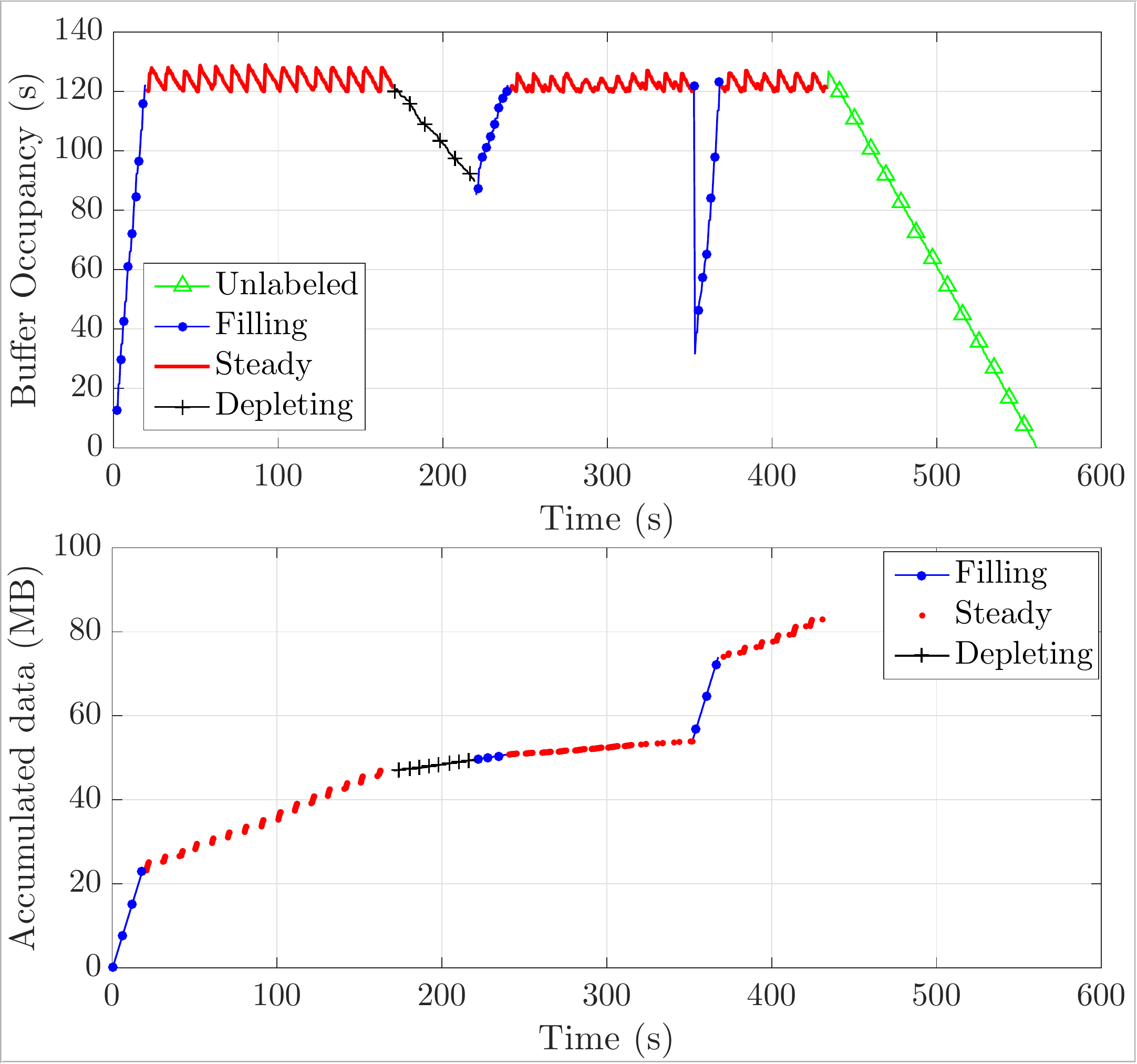} 
	\caption{Example of a labeled video flow \cite{TalkShowMovie}, using the setup in Section \ref{sec:setup} under experimental scenario (s4)} 
	\label{fig:guiExample}  
\end{figure}

\figurename{ \ref{fig:guiExample}} shows a typical HAS session, as measured using the setup of Section \ref{sec:setup}. The top figure shows the play-back buffer level in seconds, directly extracted from client's streaming application, while the bottom figure displays the accumulated streaming data over time, as recorded in our network traces. From this example, we can observe the three characteristic states of an HAS session.  
First, while the buffer is not sufficiently full, there is an initial burst of data where a new segment is requested immediately after the complete download of the previous one, leading to a streaming rate higher than the video bit-rate. We denote this period as \emph{filling state}, since the client quickly fills the buffer to a certain level, equal to $120$s in this example. Once this target is reached, the \emph{steady state} takes place where the streaming rate matches the video bit-rate, keeping the buffer level stable. This is achieved by a segment request pattern that leads to short packet bursts of one or more segments, followed by idle data transmission periods. Furthermore, due to the dynamic wireless channel conditions and the presence of diverse bottlenecks in the video delivery system, throughput may drop below the video bit-rate during streaming. We impose this case in the example of \figurename{ \ref{fig:guiExample}} by applying rate throttling in the interval $\left[170,320\right]$s. At the beginning of this interval, the client tries to download data with the available throughput, but this is not enough to support the current video bit-rate and the buffer level inevitably decreases. We define this period as a \emph{depleting state}. Then, in order to avoid a forthcoming video stall by letting the buffer run empty, the HAS policy switches to a lower video quality, in this example at $220$s, with an average bit-rate below the current throughput, leading to a second filling state and a subsequent steady state. After the end of the rate-throttling interval, higher throughput is again available and a new quality change leads to a third filling state, since the buffer is quickly filled with segments of higher quality. A last steady state takes place when the buffer target is reached again and then, after the entire video is transmitted, the session ends by playing-out the remaining bits from the buffer. We leave this part as unlabeled in the top figure, since there is no respective streaming data.  

\begin{table}[!t]
	\renewcommand{\arraystretch}{1.3}
	\caption{Recorded Parameters}
	\label{table:RecordedParameters}
	\centering
	\begin{tabular}{lll}
		\hline
		\bfseries Layer & \bfseries Name & \bfseries Description \\   \hline
		Network & $p_{\emph{srcIP}}$ & packet source IP address \\ 
		& $p_{\emph{srcPort}}$  & packet source port number \\ 
		& $p_{\emph{dstIP}}$ & packet destination IP address  \\ 
		& $p_{\emph{dstPort}}$  & packet destination port number \\ 
		& $p_{\emph{size}}$  & packet payload size (Bytes) \\ 
		& $p_{\emph{time}}$  & packet arrival time (s) \\ 
		& $p_{\emph{protocol}}$   & packet transport protocol \\ \hline
		Application & \emph{bh} & buffer level (s)\\ 
		& \emph{videoid}  &  video ID -- used for sanity check  \\
		& \emph{fmt}  & video quality (itag) \\ 
		& \emph{afmt}  &  audio quality (itag) \\ 
		& \emph{timestamp}   & time of buffer level entry (s) \\ \hline
	\end{tabular} 
\end{table}

Finally, we assume that a list of network-layer parameters can be recorded and logged to a file by observing each packet flow. This information can be observed at Layer 3 and upper Layer 2, i.e. before Radio Link Control (RLC) frame concatenation. A complete list of the required  network information at packet level is provided in Table \ref{table:RecordedParameters} and used to calculate numerical attributes, namely \emph{features}, over multiple packets of the same flow. Table \ref{table:RecordedParameters} also includes the recorded application-layer parameters from YouTube, which we collect in order to establish the ground-truth for training purposes. This allows to create a training data set, represented by matrix $\mathbf{T}\in \mathbb{R}^{N\times{\left(M+1\right)}}$, as a set of input (features) and output (label) pairs for $N$ sampling periods of duration $T_s$:    
\begin{equation}
	\mathbf{T} = \left[{
		\begin{array}{cc}
			\mathbf{x}_1 & y_1 \\
			\mathbf{x}_2 & y_2 \\
			\vdots & \vdots\\  
			\mathbf{x}_N & y_N \\
		\end{array}} 
	\right]
	\label{eq:trainingDataset} 
\end{equation}
where $\mathbf{x}_i=\left[x_{i1},x_{i2},...,x_{iM}\right]$ is a vector of $M$ features calculated at $i$-th sampling period, $y_i$ is the corresponding label value and each class is assigned a unique numerical value. This data set is then used to train a set of ML classifiers. Classification belongs to the category of supervised machine learning that involves a set of pre-classified or labeled data, associated with a set of features corresponding to this data. This input is then used to train a model by creating a set of rules in order to classify new instances based only on their features. The training phase usually requires a large training data set for better performance and may be time consuming, but can be performed off-line. Real-time classification uses the already trained  classifier in order to predict the label $y$ of the previously unseen feature vector $\mathbf{x}$ and in general can be very time efficient.    

\section{Traffic profiling}
\label{sec:TPmodel}
\begin{figure}[t]
	\centering
	\includegraphics[width=1\linewidth,trim={0.5cm 0.5cm 0.5cm 0.5cm},clip]{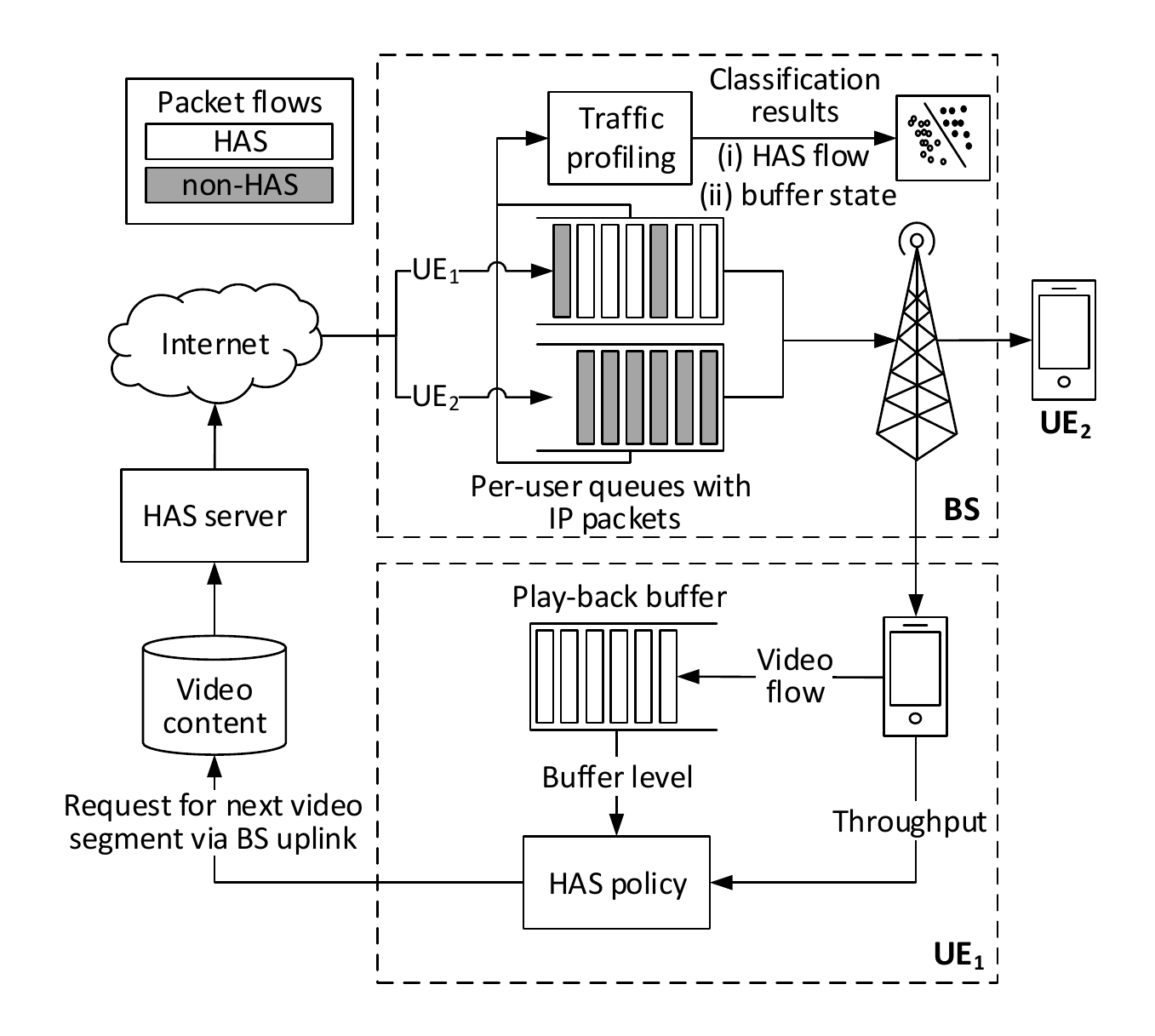} 
	\caption{Example of traffic profiling at the BS with 2 users; UE$_1$ with HAS traffic and UE$_2$ with non-HAS traffic}
	\label{fig:SystemTP}  
\end{figure}

The proposed traffic profiling system adds a module that monitors packet flows at the edge of a mobile network, i.e. at any edge-router, gateway, BS or access point. \figurename{ \ref{fig:SystemTP}} shows an application example with traffic profiling deployed at the BS monitoring the traffic of 2 User Equipments (UEs), where arriving packet flows are placed in user-specific queues and served by the BS scheduler. 

At the core of traffic profiling, information directly observable at packet level is used to construct a set of features, which are later used for the ML models. Our first classification problem is to accurately distinguish HAS flows from a set of packet flows with arbitrary traffic. Then, once an HAS packet flow is identified, we apply a second classification to detect the various buffer streaming states in real time during the HAS session. In summary, the main functionality of the proposed traffic profiling system is as follows:
\begin{enumerate}
	\item \textit{Collection of packet information:} At the transport layer, observable information at packet level is collected for each monitored packet flow, even for encrypted traffic. 
	\item \textit{Construction of features:} For each flow, features are calculated based on the collected packet information and used to build ML  models that recognize statistical properties of the flow.
	\item \textit{Detection of HAS flows:} A HAS flow is distinguished from other packet flows in real time by plugging the constructed features into the trained classifiers.
	\item \textit{Buffer state estimation for HAS flows:} Different streaming states for each HAS flow are identified in real time by using similar features and classification models. 
\end{enumerate}

\subsection{Classification}
A summary of all the defined classes is shown in Table \ref{table:Labels}. Video flow detection is formulated as a binary classification problem in our model. Each individual packet flow is classified either as `HAS' or `non-HAS' traffic, which in our case represents any measured non-HAS traffic, i.e. file downloads and Web browsing. Since we are mainly interested in video streaming, a study that attempts to distinguish a multitude of different applications with one class per application is out of the scope of this paper. 	

The problem of buffer state classification is more demanding, since there are more than two classes involved, potentially varying at any new sampling interval inside the same HAS flow. In this case we define a set of four different classes, i.e. `filling', `steady', `depleting' and `unclear', where the first three have been explained in Section \ref{sec:systemModel} and the last one is simply introduced to cover few remaining cases and mainly comprises instances where the buffer is not close to the target level, but the video bit-rate is almost equal to the available throughput, leading to a slowly varying buffer. 

\begin{table}[!t]
	\renewcommand{\arraystretch}{1.3}
	\caption{Set of defined classes}
	\label{table:Labels}
	\centering
	\begin{tabular}{lp{0.17\columnwidth}p{0.51\columnwidth}}
		\hline
		\bfseries Problem & \bfseries Name & \bfseries Description \\ 
		\hline
		Service type & \emph{HAS} & HAS traffic \\ 
		& \emph{non-HAS}   & non-HAS traffic (Web, downloads) \\ \hline
		Buffer state & \emph{filling}   & streaming rate is higher than video bit-rate, buffer is filling \\ 
		& \emph{steady}  & streaming rate matches video bit-rate, buffer target level is reached    \\
		& \emph{depleting}  & streaming rate is lower than video bit-rate, buffer level decreases \\ 
		& \emph{unclear}   & all other cases, e.g. streaming rate is close to video bit-rate but buffer target level is not reached  \\ \hline
	\end{tabular}
\end{table}

Based on our training data set, we evaluate the following five different ML classifiers, all of them well known in the ML literature:
\begin{enumerate}
	\item \emph{Support Vector Machines} (SVM): finds the best hyperplane separating data points of different classes. 
	\item \emph{$k$-Nearest Neighbors }(KNN): each sample is assigned to the most common class among its $k$ nearest neighbors.  
	\item \emph{Boosted trees}: uses AdaBoost \cite{Adaboost} algorithm to emphasize previously mis-modeled training instances. 
	\item \emph{Random Forest} (RF) \cite{RandomForest}: builds many decision trees and assigns instances to the class that most trees agree on.
	\item \emph{RUSBoost trees}: a hybrid sampling/boosting ensemble method with RUSBoost algorithm \cite{RUSboost} for skewed training data. 
\end{enumerate}

For their implementation we use the `Statistics and Machine Learning' toolbox from MathWorks \cite{MatlabML}. Specifically for SVM, a one-versus-one coding design is selected for the buffer state classification, since more than two classes are involved. Moreover, we use standardization as a rescaling method of features for classifiers that calculate the distance between two points, i.e. SVM and KNN, since our features have a wide range of values. For the evaluation of all five classifiers, we perform $k$-fold cross-validation, a common approach to train and validate a data set. This means that the training data set is randomly split into $k$ equal sized partitions. Then, each partition in turn is used as the validation data and the remaining $k-1$ are used for training. This process is repeated $k$ times  with each of the $k$ subsets used exactly once as the validation data set.

\subsection{Feature construction}
We design a small feature set in order to capture the essential information of HAS traffic at low complexity. A large number of features can often have a negative impact on the performance of ML algorithms \cite{surveyMLclassification} and also introduces higher computational and memory requirements, which are undesirable for practical implementations. The complete feature set is presented in Table \ref{table:Features}. For the purpose of our calculations we adopt a time sliding window approach, i.e. we continually measure these features every sampling period $T_s$ at $t_w=T_s, 2T_s, \dots, NT_s$ over a time window of duration $T_w=nT_s$  with $n\geq1$. Both for video flow and buffer state classification, we calculate the features of Table \ref{table:Features} over $L$ different time windows in parallel, leading to a total number of $M=5L$ features. This helps us to capture both short-term and long-term fluctuations at the cost of an increased feature space. We will revisit the impact of different windows on the classification performance when we present our results in Section \ref{sec:results}. 

The first feature in Table \ref{table:Features}, \emph{DLrate}, is simply the downlink rate of the packet flow in bit/s, given by  
\begin{equation}
	\emph{DLrate} = 8\cdot\dfrac{\sum_{p\in \mathcal{P}_{\emph{DL}}} p_{\emph{size}}}{T_w},
	\label{eq:DLrate} 
\end{equation}
where $p$ is the index of a packet and $\mathcal{P}_{\emph{DL}}$ is the set of packets with (i) $p_{\emph{time}}\in\left[t_w-T_w,t_w\right)$ and (ii) $p_{\emph{dstIP}}$ equal to the IP address of the client. This feature is particularly useful for buffer state classification, as it can reflect the difference between a filling and a steady state for similar throughput, or indicate a depleting state. 

\emph{DLrate} is complemented by our second feature
\begin{equation}
	\emph{DLload} = \dfrac{\sum_{p\in \mathcal{P}_{\emph{DL}}} {\left(\Delta p_{\emph{time}} \cdot \mathbf{1}_{\Delta p_{\emph{time}}\leq h_t}\right)}}{T_w},
	\label{eq:DLload} 
\end{equation}
where $\Delta p_{\emph{time}}$ is the IAT of two successive packets and $\mathbf{1}_{(.)}$ is the indicator function, equal to one when the IAT is less than a specified threshold $h_t<T_s$. \emph{DLload} measures the percentage of the time that is used for downlink transmission. Since it is normalized by the window duration $T_w$, $\emph{DLload} \in [0,1]$. The numerator of \eqref{eq:DLload} models the duration of continuous transmission and allows \emph{DLload} to distinguish long from short data bursts, a characteristic of different buffer states as we see in \figurename{ \ref{fig:guiExample}}. 

\begin{table}[!t]
	\renewcommand{\arraystretch}{1.3}
	\caption{Set of features calculated over each time window}
	\label{table:Features}
	\centering
	\begin{tabular}{lll}
		\hline
		\bfseries Name & \bfseries Description & \bfseries Unit\\   \hline
		\emph{DLrate}  & downlink transmit rate & bit/s  \\  
		\emph{DLload}  & fraction of used transmission time & 1\\  
		\emph{ULnPckts} & number of uplink packets & 1\\  
		\emph{ULavgSize} & average uplink packet size & Bytes\\  
		\emph{ULstdSize}  & standard deviation of uplink packet size & Bytes\\  \hline
	\end{tabular}
\end{table}

The third selected feature, denoted as \emph{ULnPckts}, is the number of uplink packets. In HAS traffic, uplink packets mostly include HTTP requests for content segments and regular ACKs. An easy way to remove ACKs from our classification is to include only packets with $p_{\emph{size}}>h_s$, where $h_s$ is a threshold for packet size, since segment requests are typically much larger. Consequently, we define 
\begin{equation}
	\emph{ULnPckts} = \sum_{p\in \mathcal{P}_{\emph{UL}}} \mathbf{1}_{p_{\emph{size}}>h_s},
	\label{eq:ULnPckts} 
\end{equation}
where $\mathcal{P}_{\emph{UL}}$ is the set of packets with (i) $p_{\emph{time}}\in \left[t_w-T_w,t_w\right)$, (ii) $p_{\emph{srcIP}}$ equal to the IP address of the client and (iii) 3-tuple $\left(p_{\emph{dstIP}}, p_{\emph{dstPort}}, p_{\emph{protocol}}\right)$ identical to  $\left(p_{\emph{srcIP}},p_{\emph{srcPort}},p_{\emph{protocol}}\right)$ of the respective downlink flow. The purpose of including \emph{ULnPckts} in the feature set is twofold. First, the packet arrival pattern for HAS traffic is different from other applications like Web browsing (less periodic) and file downloads (usually only few requests at the beginning). Secondly, the number of requests can indicate different buffer classes even for similar \emph{DLrate} and \emph{DLload}. Such a case is shown in \figurename{ \ref{fig:guiExample}}, where the second filling state is separated by the previous depleting state due to the increased number of uplink packets, since more segments are downloaded after the quality switch.

As fourth and fifth feature, \emph{ULavgSize} and \emph{ULstdSize}, we select the arithmetic mean and standard deviation of $p_{\emph{size}}$ over the previously identified \emph{ULnPckts}. Both features capture the main statistics of uplink packet size and are basically selected to improve packet flow classification. For a single video in the same streaming session, we expect a similar size for uplink packets of consecutive segment-requests. Moreover, we expect this size to vary only slightly over different videos of the same service. Such characteristic packet size distribution is not necessarily true for Web traffic, as shown in \cite{PacketSize}.

Finally, it is worth highlighting that we intentionally exclude TCP-specific features, such as TCP flag, sequence number and window. Although this type of data may be useful, as shown in \cite{Orsolic2017}, our goal is to provide a minimal set of features that is generic enough to cover both TCP and QUIC/UDP traffic. In the same spirit, we also exclude video segment-related features, such as segment size and inter-request time that are used in \cite{Dimopoulos}. Since we are designing an alternative to DPI for encrypted packet flows, we cannot assume that a reliable dissection scheme for detecting segments is in place. Identifying segments and segment requests by traffic profiling, however, would require strong assumptions on the request pattern of the OTT (e.g., to rule out cumulative requests for segments). Segment-related features are also problematic in the case of QUIC/UDP traffic, where video and audio segments are multiplexed in a single packet flow. For QUIC traffic, our experiments showed that an audio segment can be requested even before the complete download of a video segment and vice versa. Such asynchronous request pattern makes the accurate detection of segments very challenging. Thus, we believe that our feature set is more robust and more general without segment-related and TCP-specific variables.

\section{Methodology}
\label{sec:ExpMethodology}

\begin{figure}[t]
	\centering
	\includegraphics[width=1\linewidth,trim={0.5cm 0.5cm 0.5cm 0.5cm},clip]{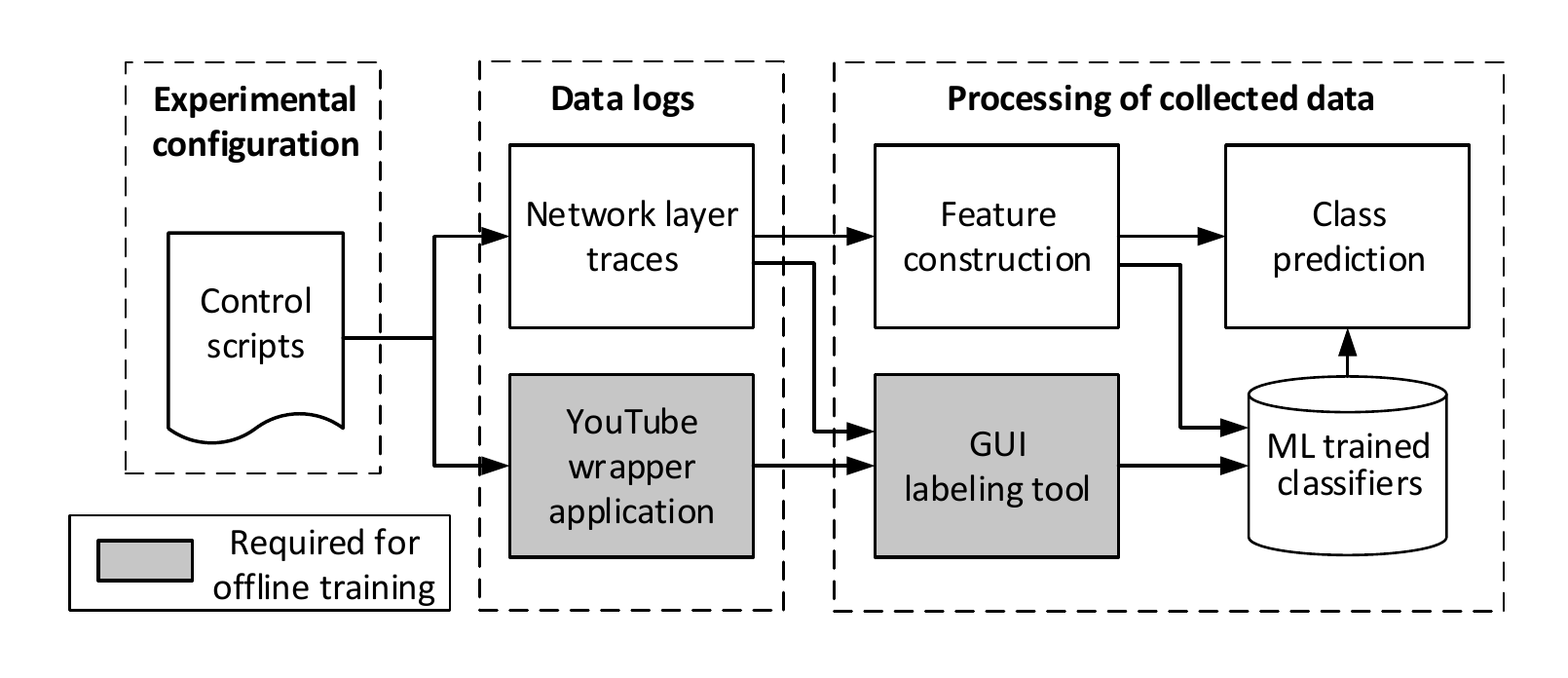} 
	\caption{Flow chart of the developed experimental setup}
	\label{fig:SystemML}  
\end{figure}

\figurename{ \ref{fig:SystemML}} provides an overview of our methodology for the collection of data and ground truth. First, a set of control scripts that run at an intermediate node drives the whole process and allows us to control the phones and configure the measurement campaign by setting the following list of parameters:
\begin{itemize}
	\item target video ID
	\item video quality profile   
	\item wireless channel profile
	\item number of measurement iterations
	\item set of network-layer data to be recorded
	\item set of application-layer data to be recorded  
\end{itemize}
As a result of the control scripts, a batch of data log files is created as soon as a video streaming session is completed. Network-layer logs are provided by a packet analyzer tool and include all the network parameters of Table \ref{table:RecordedParameters}.   

\subsection{Ground truth and labeling process}
\label{sec:labeling}
Before we proceed with labeling, we need to obtain certain parameters from the YouTube application. For this reason, we developed a \emph{wrapper application} for YouTube and installed it in the client's device, allowing control and automatic interaction with YouTube's Android interface. Application wrapping is the process of applying a management layer to a mobile application without changing the underlying application. This way, video ID selection, quality adjustment and settings configuration, such as disabling auto-play or enabling YouTube video statistics, can be programmed and executed without any user intervention. Moreover, the wrapper application allows to monitor the progress of the streaming session and retrieve application-layer information, as listed in Table \ref{table:RecordedParameters}, that is available through the \emph{stats for nerds} option in YouTube's interface. More details regarding the available information through the statistics module of YouTube can be found in \cite{Candid}. These application parameters are recorded in a log-file twice per second by using a \emph{clipboard application}, which copies debug information from the data buffer (clipboard) of the phone directly into a file. All commands required for the implementation of an experimental scenario are transferred to the device via the Android Debug Bridge (ADB) tool, as soon as they are parsed from the respective configuration script.

The process of labeling is trivial for the binary classification problem of HAS flow detection, as we simply need to assign the same label for all sampling periods of a packet flow. By isolating experiments with HAS traffic, it is easy to detect HAS packet flows, by checking the total packet size of each flow, as only streaming flows should have significant size. However, in our approach we also visually verify labels by comparing the data pattern of a flow to the pattern of the buffer level obtained by the application layer. 

On the other hand, labeling for buffer state classification is challenging,  since there is no ground truth information about buffer states available from the application layer and one has to rely on the  buffer level logs. An automated procedure based on algorithms that analyze these logs is possible, but requires careful design covering many different patterns and outliers that could otherwise lead to false results. Instead, in our methodology we decided to introduce a process based on manual inspection. For this purpose, we developed a Graphical User Interface (GUI) tool that loads parsed network and application logs, provides both buffer level and accumulated data plots as in \figurename{ \ref{fig:guiExample}}, and additionally allows a user to manually:
\begin{itemize}
	\item select for which packet flow (from the list of captured flows) to plot the respective accumulated data
	\item specify disjoint time intervals for which a unique label can be assigned
	\item select a label (from the list of available labels) to associate with a previously defined interval 
\end{itemize}
Besides labeling, such a tool is also useful in order to verify that an experiment is successfully completed and properly recorded, get insights about the HAS policy and investigate experiments with unusual client behavior.

\subsection{Experimental setup}
\label{sec:setup}

\begin{figure}[!t]
	\centering
	\includegraphics[width=0.9\linewidth,trim={0cm 0cm 0cm 0cm},clip]{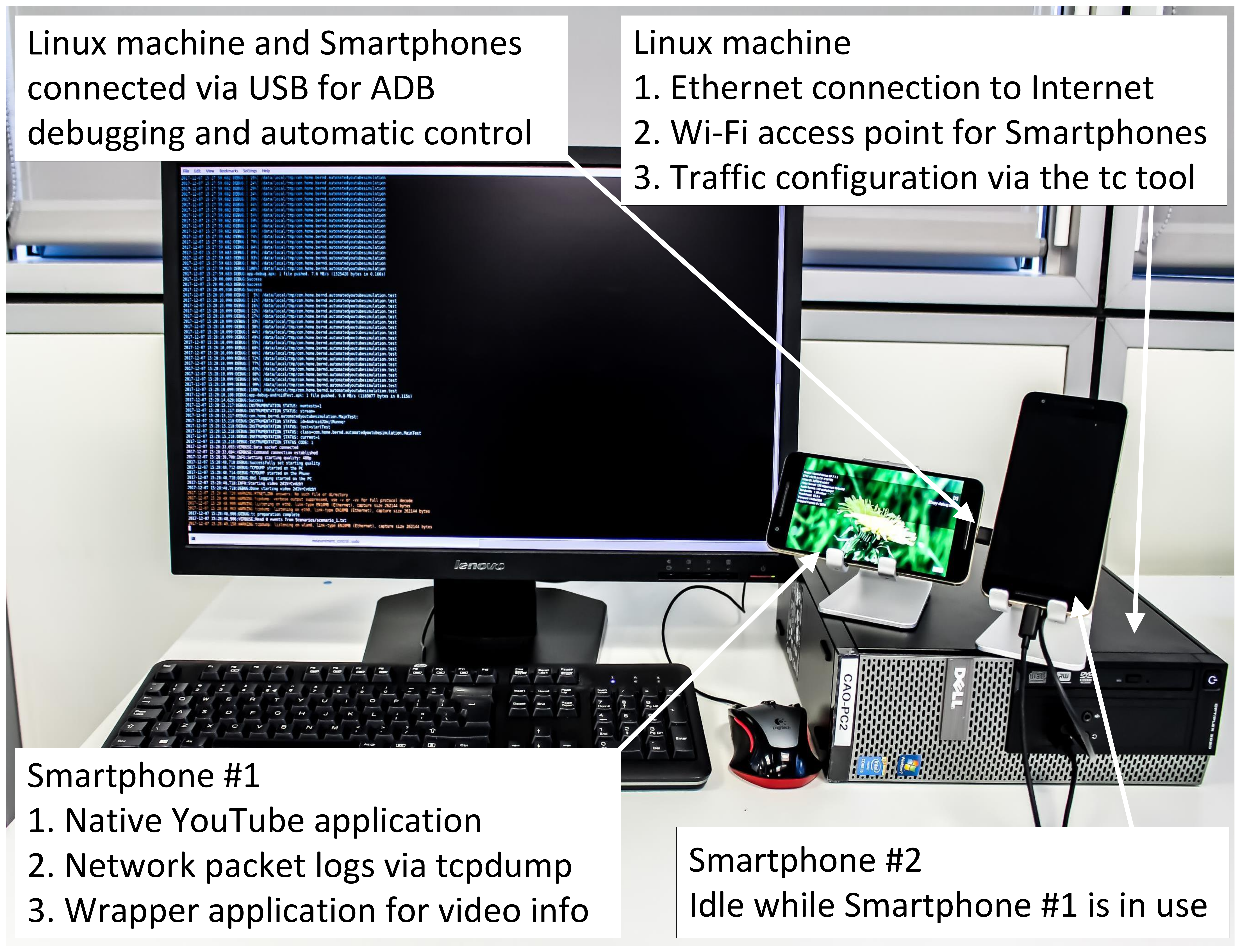} 	
	\caption{Setup for measurements and ground truth}
	\label{fig:ExpSetupPhoto}  
\end{figure}

The testbed shown in \figurename{ \ref{fig:ExpSetupPhoto}} is designed in order to measure YouTube traffic in an automatic, controlled and reproducible manner. Two Android Smartphones (Huawei Nexus 6P, baseband version: angler-03.78, Android 7.1.1 with the security patch from December 5\textsuperscript{th}, 2016) are connected via a Wireless Local Area Network (WLAN) to a Linux computer (Kernel 3.16.0-71-lowlatency) that operates as a WLAN access point. The computer is connected to the Internet via a T1 line, that acts as a gateway for the Smartphones, and controls the phones via a Universal Serial Bus (USB) connection. The WLAN operates in IEEE 802.11g mode at a carrier frequency of 2412 MHz.

Controlled configuration of network parameters such as rate, delay and Packet Error Rate (PER), extends the reproducibility and increases the functionality of our testbed. The WLAN interface,   combined with the traffic configuration (tc) tool provided in the Linux kernel \cite{tc}, allow to configure network traffic parameters and emulate sufficiently the networking dynamics of Long-Term Evolution (LTE). Network layer packet logs are recorded with tcpdump \cite{tcpdump} on both the computer and the Smartphones. The traffic is generated with the native YouTube application (version: 12.32.60) for Android, which according to our observations, performs standard DASH operation \cite{DASH}. The YouTube application protects its streaming traffic via TLS encryption and consequently, HTTP queries are sent to the server TCP port 443. Over the course of our measurements, the QUIC protocol was used in most cases.

\subsection{HAS scenarios}
\label{sec:Scenarios}	

\begin{figure}[!t]
	\centering
	\includegraphics[width=0.85\linewidth,trim={0.7cm 0.1cm 1cm 0.6cm},clip]{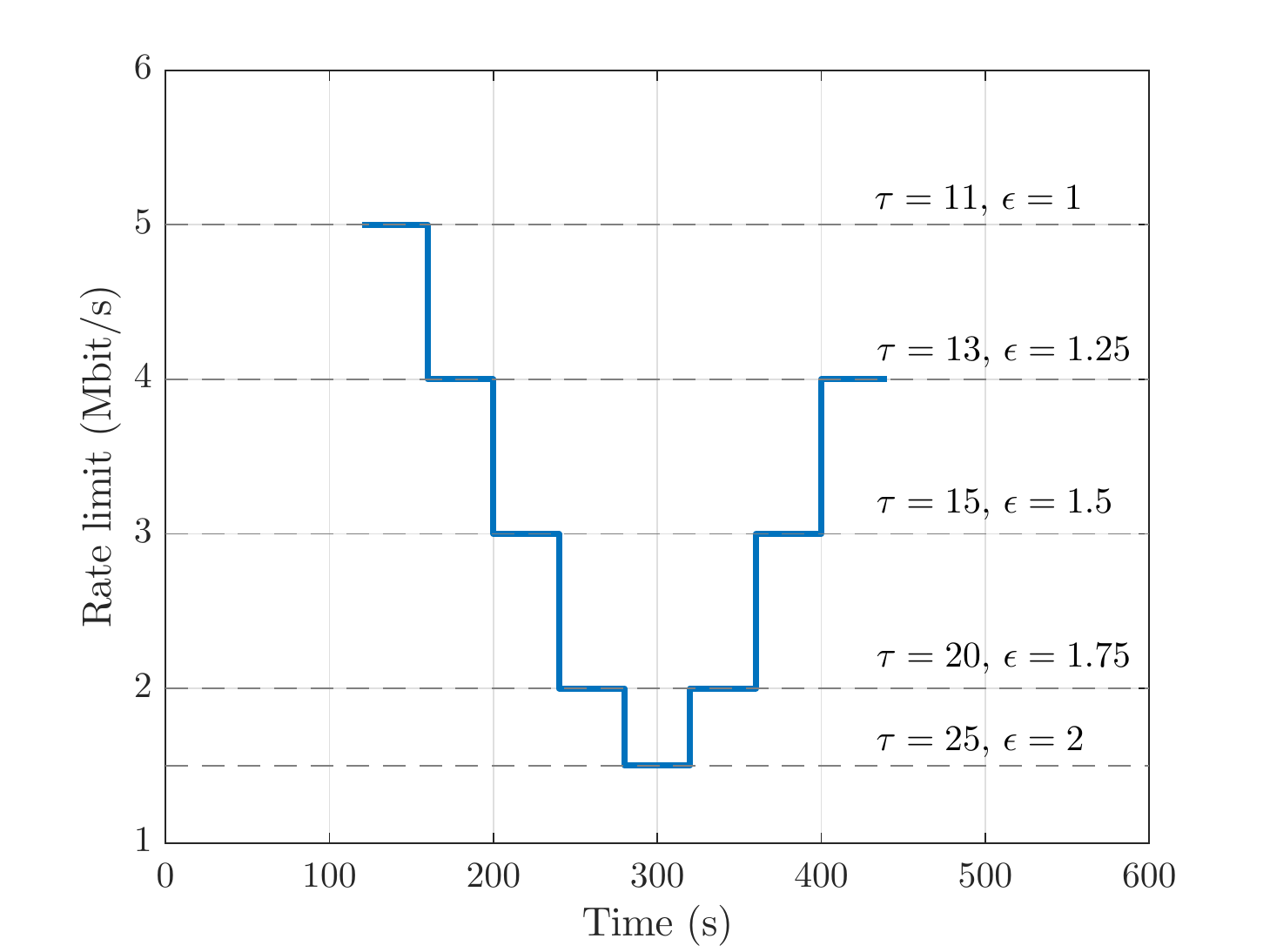} 
	\caption{Traffic configuration of rate (Mbit/s), delay $\tau$ (ms) and PER $\epsilon$ (\%) for (s6) according to DASH-IF guidelines}
	\label{fig:DASH-IF}  
\end{figure}

In order to cover a variety of representative streaming situations, we design and include in our experiments 8 different scenarios, as specified in Table \ref{table:scenarios} and listed below: 
\begin{itemize}
	\item[(s1)] Medium quality (480p), no adaptation and  no traffic configuration for the entire video.
	\item[(s2)] High quality (720p), no adaptation, no traffic configuration for the entire video.
	\item[(s3)] Quality Change (720p to 480p) at a random time in the interval $[120,240]$s, no adaptation and no traffic configuration.
	\item[(s4)] Adaptive quality, rate limitation at 500kbit/s starting randomly in the interval $[120,240]$s with a duration of 150s. 
	\item[(s5)] Adaptive quality with constant rate limitation at 1024kbit/s for the entire video.
	\item[(s6)] Adaptive quality based on DASH Industry Forum (DASH-IF) implementation guidelines \cite[Table 5]{Forum2014}. Besides the first step at 120s, all the next steps are applied every 40s. Rate, delay and PER traffic configurations are illustrated in \figurename{ \ref{fig:DASH-IF}}.
	\item[(s7)] Adaptive quality, rate limitation at 120s by switching from 3Mbit/s to 100kbit/s and back to 3Mbit/s every 40s and 45s respectively until the end of the video.
	\item[(s8)] Adaptive quality, rate limitation of 100kbit/s starting at 120s and 400s, with a duration of 60s in both cases. 
\end{itemize}

In order to improve the quality of our training dataset, during the experimental design we made sure that our scenarios depict at least some clear state transitions. For this reason, we decided to leave the first 120s of each scenario without rate limits, besides (s5), since we verified from our experimental results that for the selected video content, a first state transition from filling to steady state appears during this time in normal streaming conditions. It should be noted that after the video starts playing, we assume that the user does not interrupt the video play-back by pausing or skipping forwards or backwards. Even though these events can  be detected, we neglect them for simplicity. 

\begin{table}[!t]
	\renewcommand{\arraystretch}{1.3}
	\caption{HAS scenarios}
	\label{table:scenarios}
	\centering
	\begin{tabular}{ccc}
		\hline
		\bfseries Scenario & \bfseries Quality & \bfseries Rate limit (kbit/s)\\ \hline
		1 &  480p & Inf \\
		2 & 720p & Inf \\
		3 & 720p$\xrightarrow{t_0}$480p & Inf \\
		&  $t_0 \in \left[120,240\right]$s  &  \\
		4 & Auto & Inf $\xrightarrow{t_0}$ 500 $\xrightarrow{t_1}$ Inf \\
		&    &  $t_0 \in \left[120,240\right]$s, $t_1=t_0+150$\\
		5 & Auto & 1024 \\
		6 & Auto & (see \figurename{ \ref{fig:DASH-IF}}) \\
		7 & Auto & Inf $\xrightarrow{\text{120s}}$ 3000  $\overset{t_0}{\underset{t_1}\rightleftarrows}$ 100   \\
		&    &  $t_0=160+85n,  n \in \mathbb{N}$\\
		&    &  $t_1=205+85n, n \in \mathbb{N}$\\
		8 & Auto & Inf $\overset{t_0}{\underset{t_1}\rightleftarrows}$ 100  \\
		&    &  $t_0\in\{120,300\}$s, $t_1\in\{180,380\}$s\\
		\hline
	\end{tabular}
\end{table}

\begin{figure*}[!t]
	\centering
	\includegraphics[width=1\linewidth,trim={0.1cm 0cm 0cm 0cm},clip]{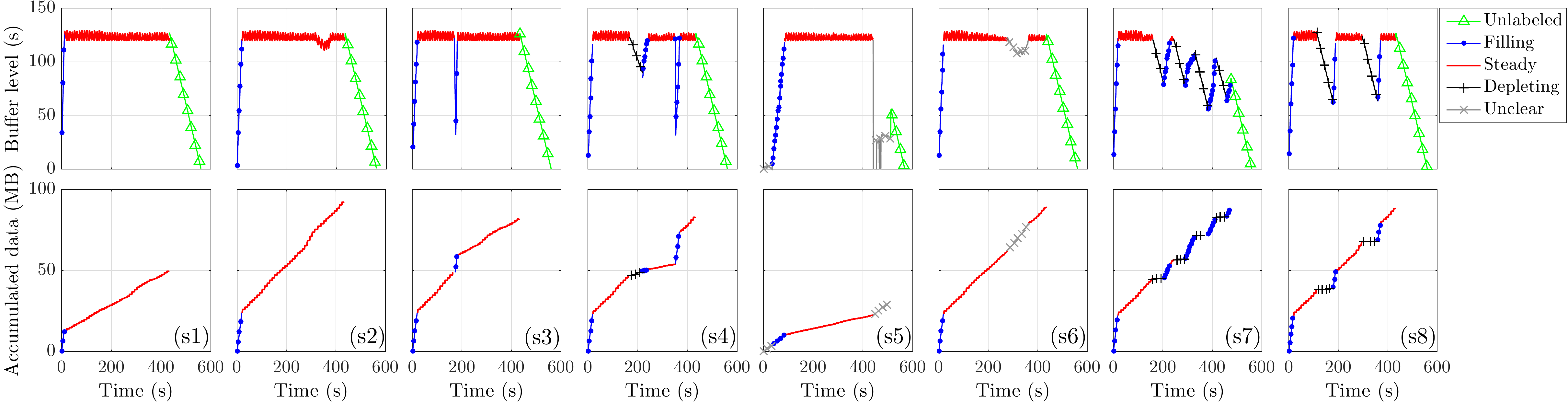} 
	\caption{Examples of buffer level and streaming data per scenario for video \cite{TalkShowMovie}, labeled according to Section \ref{sec:labeling}}
	\label{fig:ScenariosAll}  
\end{figure*}

Specifically, scenarios (s1) and (s2) are chosen in order to study the performance of the algorithm in simple cases with a single filling and steady state, for two quality levels with significant bit-rate difference. Then, (s3) is selected to verify that the algorithm can also detect multiple transitions between streaming states throughout the video session. (s4) is a more challenging scenario due to the introduced buffer depleting state as a result of rate throttling. The randomness in (s3)-(s4) is introduced in order to decrease the correlation with the video encoding distribution. The main reason behind the choice of (s5) is to include state changes under a rate that is common for 3G networks. In (s6) we emulate more complex streaming conditions where the possibility of unclear buffer states is higher. \figurename{ \ref{fig:DASH-IF}} shows the traffic configuration of this scenario, where a high-low-high rate profile is used. The indicated values of delay $\tau$ refer to transmission delay at the access point and should not be confused with the Round Trip Time (RTT), which in our experiments had a mean value of 30ms. Finally, (s7)-(s8) are added to cover poor connection cases where streaming is not supported even with the lowest available video quality. A representative example for each scenario is shown in \figurename{ \ref{fig:ScenariosAll}}, where both buffer level and accumulated streaming data are labeled according to Section \ref{sec:labeling}.

For the selection of streaming content, our target is to capture a variety of video bit-rate distributions amongst characteristic video clip types. To this end, we make a selection of 3 movies, as we regard them to be representative of different typical video content types. As a first choice, we study Tears of Steel (\emph{TOS}) \cite{TearsOfSteelMovie}, a high motion semi-animated open action movie which is commonly used for testing video codecs and streaming protocols and that is also recommended in the measurement guidelines of DASH-IF \cite{Forum2014}. TOS is of 12:14 min duration and represents a high motion video clip. We also select a nature documentary (\emph{Nature}) \cite{NatureMovie} of 9:21 min duration that contains complex scenes with gradual changes. As a third choice, a talk-show (\emph{TalkShow}) \cite{TalkShowMovie} of 9:19 min duration is selected as streaming content of a low-motion video clip. All clips are encoded with the H.264 codec in an MP4 container. Our selection of the clips intentionally excludes monetized content, as we want to avoid advertisements at the beginning or during the video session interfering with our measurements. This is done without loss of generality, as advertisements can be identified as different packet flows and separated from the rest of the streaming data. Table \ref{table:movies} presents the YouTube video id of each tested movie along with main characteristics, i.e. available representation range, duration and frame rate (fps).

\begin{table}[!t]
	\renewcommand{\arraystretch}{1.3}
	\caption{Streaming content}
	\label{table:movies}
	\centering
	\begin{tabular}{ccccc}
		\hline
		\bfseries Video & \bfseries id & \bfseries Quality  & \bfseries Duration & \bfseries fps \\
		\hline
		TOS & OHOpb2fS-cM & 144p-1080p & 12:14 min & 24 \\ 				
		Nature & 2d1VrCvdzbY & 144p-1080p & 09:21 min & 30  \\
		TalkShow & N2sCbtodGMI & 144p-1080p & 09:19 min & 25 \\         
		\hline
	\end{tabular}
\end{table}

\subsection{Data sets}
We test each of the 3 movies across 8 different scenarios, with 30 repetitions per scenario-movie tuple. In total we gathered 720 measurements, which were conducted in Paris, France, over the course of two weeks in October 2017. To assure full batteries, both phones measured in alternating cycles of 12 hours. By measuring on a continuous basis, we capture potential traffic congestion during overloaded periods, but also traffic at normal or low load. This is done to avoid a statistical bias if only certain periods within a day are used for measurements. Before each measurement the cache memory of the YouTube application is cleared to ensure that no pre-fetched part of the video is saved in the device. 

For the purpose of video flow classification, we also conducted measurements of non-HAS traffic, by including 2 additional scenarios, i.e. file download and Web browsing, and included them in our training dataset. In practice, this type of mixed traffic reflects the presence of multiple users sharing the network at the same time, as the current version of native YouTube application does not allow multitasking. However, such a limitation can easily change in the future, e.g. users simultaneously streaming a YouTube video and accessing the Web. Note that YouTube's RED subscription-based service \cite{youtubered} already provides such an option, by allowing to stream in the background while the YouTube application is minimized.

\begin{table}[!t]
	\renewcommand{\arraystretch}{1.3}
	\caption{Number of samples per class in the training dataset}
	\label{table:nSamples}
	\centering
	\begin{tabular}{cccccc} 
		\hline
		\bfseries Class  & filling & steady & depleting & unclear & \bfseries Total  \\ \hline
		HAS &  23296 & 82225 & 9722 & 24297 & 139540 \\
		& (16.7\%) & (58.9\%) & (7.0\%) & (17.4\%) & (100.0\%) \\
		non-HAS & --  & -- & -- & -- &  8071  \\ \hline
	\end{tabular}
\end{table}	

Based on our experimental testbed, downloads of generic data files of size 100 MB, 200MB and 512 MB are performed by accessing a host server, i.e. \url{https://www.thinkbroadband.com/download} in our case, via the built-in Web browser of the Smartphone.  For Web browsing, we developed an automatic framework that navigates to a specified Web page from a list of pages via the built-in Web browser of the Smartphone and then proceeds to the next page in the list after a random amount of time of $\left[5,15\right]$s, emulating users browsing on the Web. In order to generate traffic with no trivial data size, we selected a list of pages that host a significant number of photos besides plain text. We explicitly exclude Web pages that host video elements and leave the problem of classifying different types of video for our future work. Similarly to the video experiments, before every experiment the cache memory is cleared in order to avoid any caching issues for subsequent measurements.

Table \ref{table:nSamples} summarizes the number of samples per class in our training dataset, collected for a sampling period of $T_s=1$s and manually labeled according to Section \ref{sec:labeling}.  Since our target is to classify transmitted data per sampling period, the values in Table \ref{table:nSamples} count only non-empty samples, i.e. seconds with one or more packets. This filtering allows us to reduce the dataset size by keeping only meaningful entries. As previously explained, this dataset is used for both  training and testing by applying $k$-fold cross-validation.

\section{Results}
\label{sec:results}

The measurement results for the proposed classification of flow type and buffer state are presented in this section. The main free parameters related to both feature construction and configuration of the ML classifiers are shown in Table \ref{table:FreeParameters}. The kernel type for SVM and the number of nearest neighbors for KNN are selected by keeping the best option in terms of overall accuracy after studying a set of commonly used values with our dataset. For the 3 methods based on trees, a small number of trees is used initially, as we want to keep low complexity and memory requirements, but we detail more on the impact of this parameter at the end of this section.  

\subsection{Video flow classification}
\label{sec:videoFlow}
\begin{figure}[t]
	\centering
	\includegraphics[width=1\linewidth,trim={0.2cm 0.1cm 1cm 0.6cm},clip]{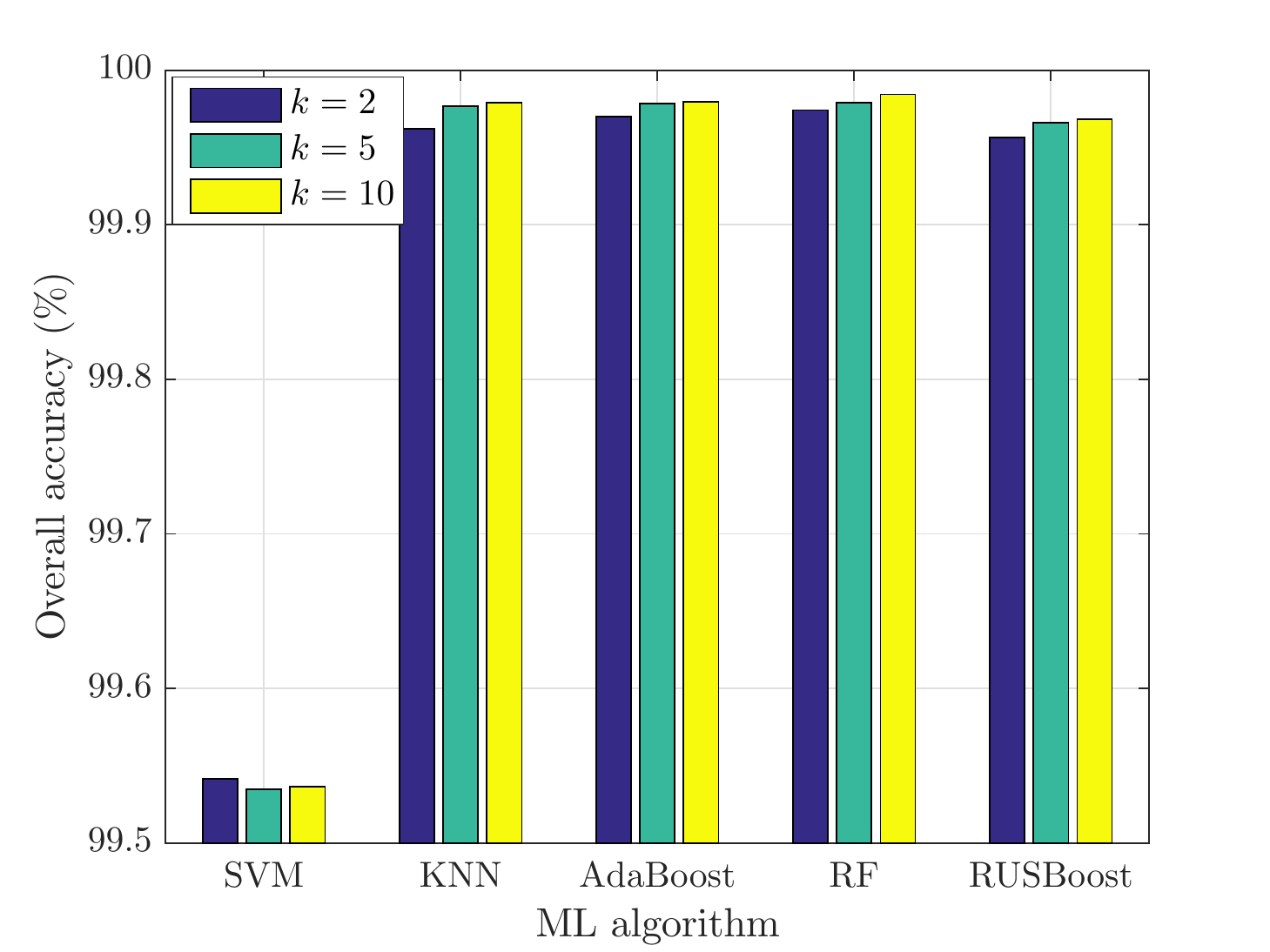} 
	\caption{Overall accuracy for flow classification by $k$-fold cross-validation}
	\label{fig:FlowID_accuracy}  
\end{figure}

\figurename{ \ref{fig:FlowID_accuracy}} presents the overall accuracy of each ML algorithm for the problem of flow classification. Overall accuracy is defined here as the ratio of correctly classified samples over the total number of samples. The performance evaluation is based on $k$-fold cross-validation, by studying 3 different values of $k$. From this figure we can easily observe that all ML classifiers perform with very high accuracy. Besides SVM that has an accuracy slightly more than 99.5\%, all the rest of the algorithms perform similarly, while RF is the best with an accuracy close to 99.98\% for $k=10$. We also notice that factor $k$ does not have a significant impact on the results, which is expected as the impact of $k$ diminishes with large data sets, verifying that the size of our dataset is sufficient to get statistically good estimates of classification performance. Thus, for the rest of the analysis we present results only for 10-fold cross-validation, which is the most commonly used setting in the literature. 

\begin{table}[!t]
	\renewcommand{\arraystretch}{1.3}
	\caption{Main parameters for features and ML classifiers}
	\label{table:FreeParameters}
	\centering
	\resizebox{1.0\linewidth}{!}{
		\begin{tabular}{lcc} 
			\hline
			\bfseries  Description   & \bfseries Value & \bfseries Unit  \\ \hline
			sampling period $\left(T_s\right)$ & 1 & s \\
			sliding window duration $\left(T_w\right)$ & $\{1,5,10,20\}$  & s \\
			threshold for \emph{DLload}  $\left(h_t\right)$   &  0.1 & s\\
			threshold for \emph{ULnPckts} $\left(h_s\right)$ & 100  & Bytes \\ \hline
			$k$-fold cross validation $\left(k\right)$ [all ML alg.]& $\{2,5,10\}$ & 1  \\
			kernel type [SVM]  & linear & 1 \\
			number of nearest neighbors [KNN]  & 1 & 1 \\
			number of trees [AdaBoost, RF, RUSBoost]  &  30 & 1 \\ \hline
		\end{tabular}}
\end{table}	
	
A confusion matrix that summarizes the results of each algorithm is presented in Table \ref{table:ConfusionMatrix_Flow}, where we highlight the performance of RF as the best algorithm according to our studies. From this table we see that HAS traffic is almost always classified correctly, while the overall accuracy is mostly affected by \emph{false positives}, i.e. samples of non-HAS traffic incorrectly classified as HAS traffic. However, the percentage of false positives is very low with a value less than 0.4\% for all algorithms apart from a 5.7\% for SVM. As a general remark, one should keep in mind that all classification results are per sampling period. In practice, a packet flow cannot change class until its termination, since it either belongs to HAS or non-HAS traffic. This fact enables post-processing methods to perform a second step of flow classification based on the existence of a dominant class, where isolated samples with predicted class different from the dominant predicted class can be neglected. 
	
\begin{table}[!t]
	\renewcommand{\arraystretch}{1.3}
	\caption{Confusion matrix for flow classification ($k=10$)}
	\label{table:ConfusionMatrix_Flow}
	\centering
	\begin{tabular}{lll} 
		\hline
		\bfseries True  & \multicolumn{2}{l}{\bfseries Predicted (SVM/KNN) } \\ \hline
		(\%) & HAS & non-HAS \\
		HAS &  99.8/100.0   &   0.2/0.0  \\
		non-HAS &  5.7/0.3  & 94.3/99.7  \\ \hline
		\bfseries True & \multicolumn{2}{l}{\bfseries Predicted (AdaBoost/RF/RUSBoost) } \\ \hline
		(\%) & HAS & non-HAS \\ 
		HAS & 100.0/\textbf{\color{red}100.0}/100.0  & 0.0/0.0/0.0\\
		non-HAS & 0.3/0.2/0.4 & 99.7/\textbf{\color{red}99.8}/99.6\\ \hline			
	\end{tabular}
\end{table}	
	
\begin{figure}[!t]
	\centering
	\includegraphics[width=1\linewidth,trim={0.2cm 0.1cm 1cm 0.6cm},clip]{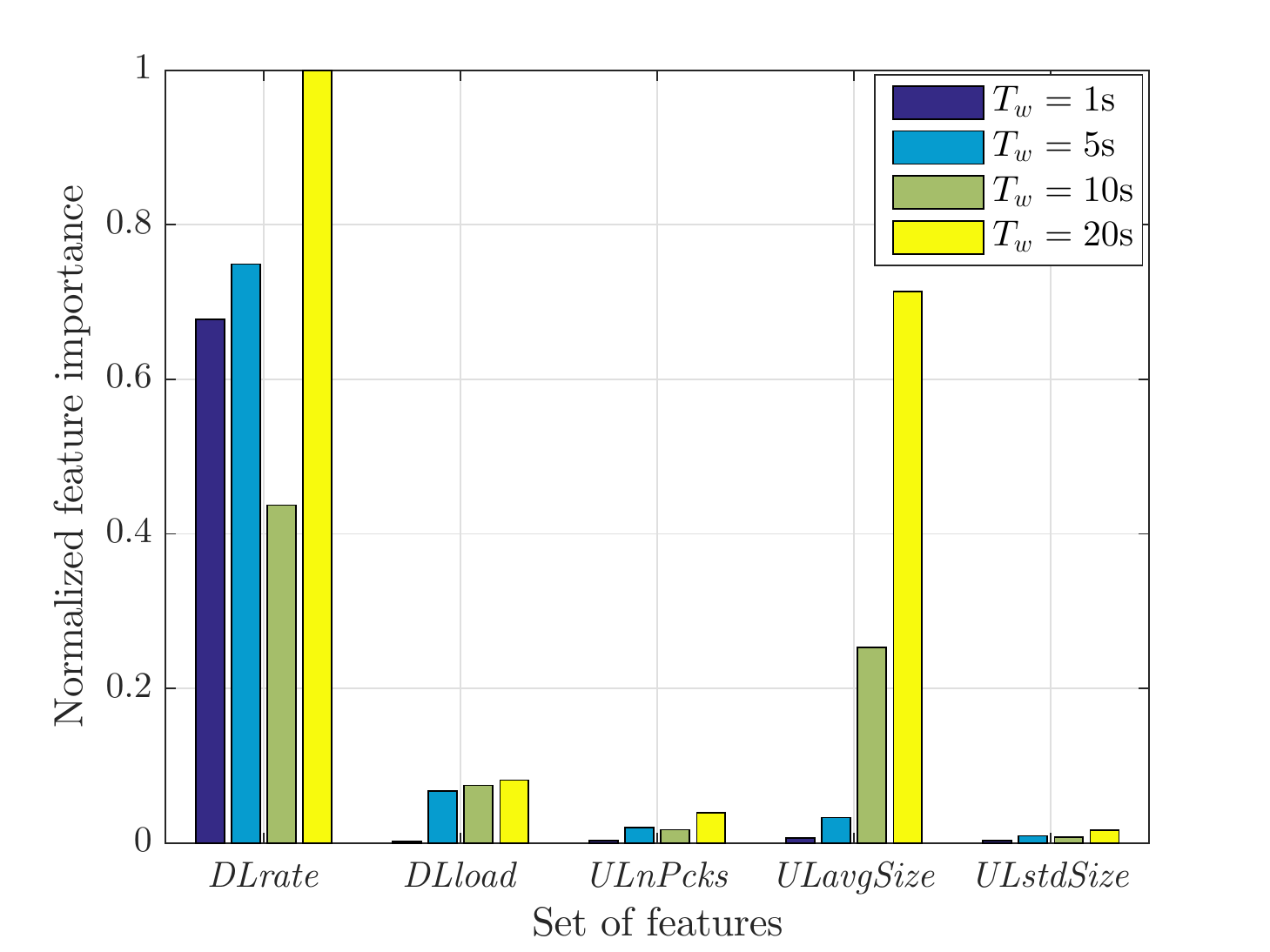} 
	\caption{Feature importance for flow classification using RF and 10-fold cross-validation}
	\label{fig:FlowID_importance}  
\end{figure}

\figurename{ \ref{fig:FlowID_importance}} shows the importance of each feature for the RF algorithm. This metric naturally ranks features according to their relevance for the classification \cite{RandomForest}. The score is normalized over the maximum obtained value and defined for a feature $m$ as the mean difference in \emph{out-of-bag error} between the original forest and a modified version where the values of feature $m$ are randomly permuted. Out-of-bag error is defined as the mean classification error over each training sample $\mathbf{x}_i$, using the votes only from trees that do not contain $\mathbf{x}_i$. Going back to \figurename{ \ref{fig:FlowID_importance}}, the complete set of $M=20$ features is included by applying $L=4$ sliding windows in parallel.  From this figure we observe that \emph{DLrate} for all windows, along with \emph{DLload} for $T_w=\{5,10,20\}$s and \emph{ULavgSize} for $T_w=\{10,20\}$s, are ranked as the most important features. First, the combination of \emph{DLrate} and \emph{DLload} for different $T_w$ allows to capture the unique HAS on-off pattern at steady state, but also to detect controlled rate changes, due to transitions between filling and steady states. This pattern is present neither in file downloads nor in Web browsing.  Moreover, statistics over the uplink packet size are also useful, as the importance value for \emph{ULavgSize} suggests. As expected, HAS requests have similar size during the entire video session that does not vary significantly over our video set, which proved to be different from what happens in our non-HAS dataset. It is worth noting that for \emph{ULavgSize}, a large window is required in order to cover the inter-request times in the steady state. Finally and perhaps counter-intuitively, \emph{ULstdSize} shows to have a negligible impact on the RF performance. Nevertheless, we keep it in our feature set as we believe that it can boost the classification performance if other types of non-HAS traffic are studied and added in the training dataset. 
	
Table \ref{table:CompTimeFlow} presents the runtime of the 5 studied ML algorithms, measured on an Intel Xeon CPU E5-4627 v2, running 32 cores at 3.30GHz with 512 GB RAM. The training time is measured for the entire dataset, while the prediction time is measured per 1000 samples. The computational time statistics are calculated over 100 repetitions, both for training and prediction. These results clearly show that SVM has a moderate prediction time and is by far the most demanding in terms of training time. On the contrary, KNN has a trivial training phase as it simply stores samples. It is, however, quite expensive in terms of prediction time since it requires to compute the distance between a new sample and all training samples. RF is slightly slower than AdaBoost and RUSBoost for the same number of trees since it builds deeper trees. Nevertheless, these 3 methods show the same behavior with fast prediction and training phases. As a general note, while fast off-line training is desirable for tuning the algorithms and updating the dataset, the critical part for a practical implementation is the runtime during the on-line phase of prediction.

\begin{table}[!t]
	\renewcommand{\arraystretch}{1.3}
	\caption{Runtime  for video flow classification (mean $\mu$ and standard deviation $\sigma$)}  
	\label{table:CompTimeFlow}
	\centering
	\begin{tabular}{p{0.17\columnwidth}p{0.12\columnwidth}p{0.12\columnwidth}p{0.12\columnwidth}p{0.12\columnwidth}} 
		\hline
		\bfseries ML alg.  & \multicolumn{2}{l}{\bfseries Training} & \multicolumn{2}{l}{\bfseries Prediction}  \\ 
		& $\mu$ (s) & $\sigma$ (s) & $\mu$ (ms) & $\sigma$ (ms) \\ \hline
		SVM 		& 463.23 & 8.08 & 59.72 & 0.24 \\
		KNN 		& 0.64 & 0.01 & 190.80 & 3.27 \\
		AdaBoost 	& 8.63 & 0.07 & 11.20 & 0.33\\
		RF 			& 10.40 & 0.34 & 11.76 & 0.23 \\
		RUSBoost 	& 3.56 & 0.10 & 10.28 & 0.32 \\	\hline
	\end{tabular}
\end{table}
	
\begin{table*}[!t]
	\renewcommand{\arraystretch}{1.3}
	\caption{Confusion matrix for buffer state classification ($k=10$)}
	\label{table:ConfusionMatrix_Buffer}
	\centering
	\begin{tabular}{lllll lllll} 
		\hline
		\bfseries True  & \multicolumn{4}{l}{\bfseries Predicted (SVM/KNN) }  & & \multicolumn{4}{l}{\bfseries Predicted (AdaBoost/RF/RUSBoost)} \\ \cline{1-10}
		(\%) & filling & steady & depleting & unclear &  & filling & steady & depleting & unclear \\ 
		filling   & 90.4/98.3 & 3.5/0.6  &  0.9/0.7 & 5.3/0.3 &  & 93.3/\textbf{\color{red}98.7}/94.2 & 4.9/0.5/2.4 & 0.5/0.6/1.7 & 1.3/0.2/1.7 \\ 
		steady     &  0.7/0.2 & 98.1/99.5 &  0.4/0.1 & 0.8/0.2 &  & 0.3/0.1/0.9 & 98.8/\textbf{\color{red}99.7}/94.5 & 0.2/0.1/2.5 & 0.8/0.1/2.1\\
		depleting & 1.2/2.1  & 13.6/1.4  & 83.0/96.4 & 2.2/0.2 &  & 2.0/1.4/1.7 & 11.0/0.9/2.8 & 86.5/\textbf{\color{red}97.6}/94.5 & 0.6/0.1/1.0\\ 
		unclear  & 4.2/0.3  & 5.6/0.6  & 0.4/0.0  & 89.8/99.0  &  & 3.3/0.2/4.8  & 4.5/0.5/3.3  & 0.3/0.0/0.7  & 91.8/\textbf{\color{red}99.3}/91.1 \\  \cline{1-10}
	\end{tabular}
\end{table*}
	
\subsection{Buffer state classification}
\begin{figure}[t]
	\centering
	\includegraphics[width=1\linewidth,trim={0.2cm 0.1cm 1cm 0.6cm},clip]{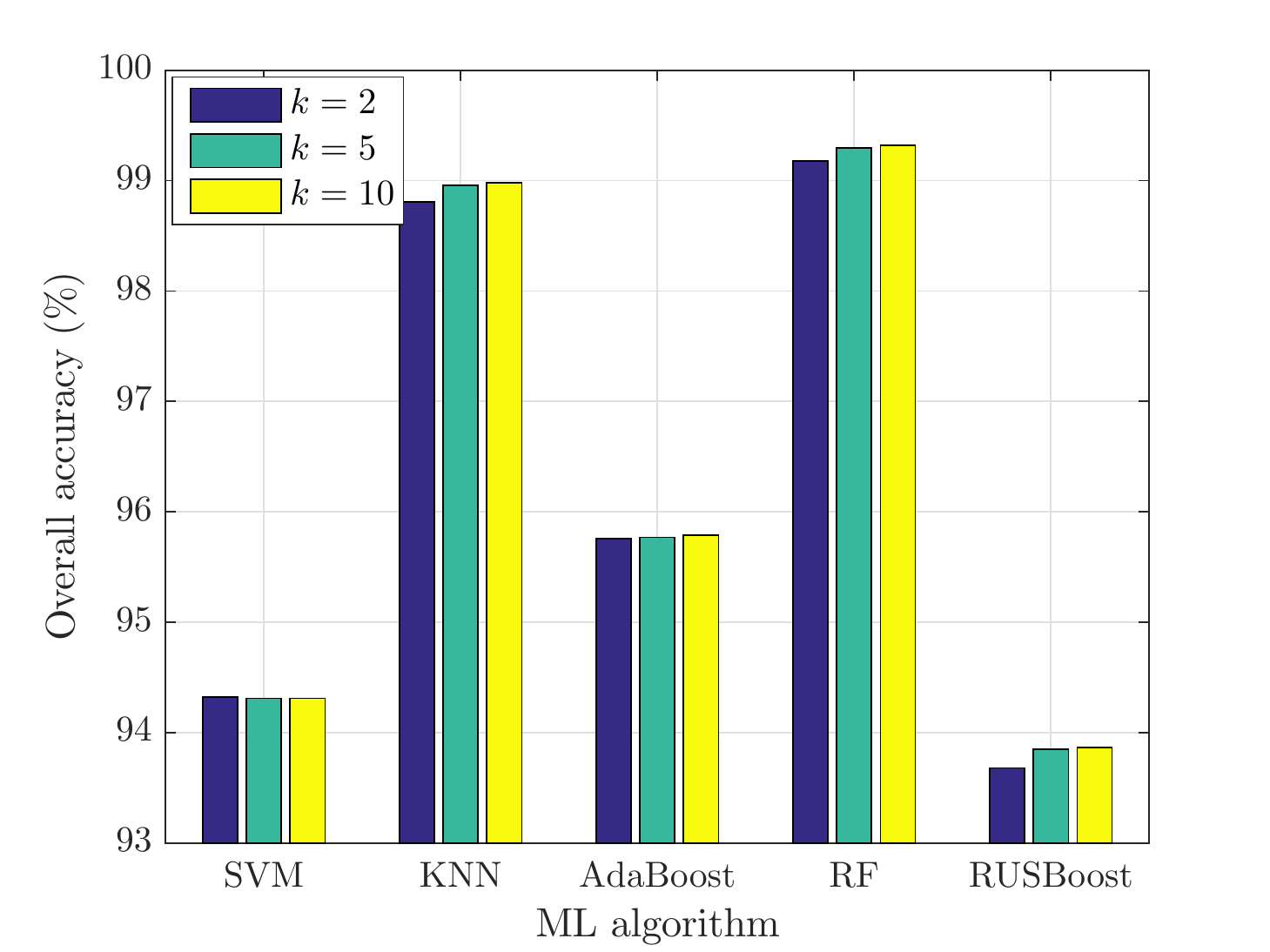} 
	\caption{Overall accuracy for buffer state classification by $k$-fold cross-validation}
	\label{fig:BufferStates_accuracy}  
\end{figure}

Having confirmed the high accuracy of HAS flow classification, we proceed with studying the performance of buffer state classification. As before, we start by presenting the overall accuracy of each ML algorithm in \figurename{ \ref{fig:BufferStates_accuracy}}, for different $k$-fold cross-validation modes. From this figure we verify once again that factor $k$ does not affect the results and thus, we keep the value $k=10$ for the rest of the presented results. RF still has the best performance with an accuracy of 99.3\% for $k=10$, while KNN follows closely with 99\% and the rest of the algorithms fall behind with a value ranging from 93.9\% to 95.8\%. A relationship between RF and KNN that may explain their similar behavior is discussed in \cite{KNN_RF_relation}, where it is shown that both can be viewed as weighted neighborhoods schemes.  
	
\begin{figure}[!t]
	\centering
	\includegraphics[width=1\linewidth,trim={0.2cm 0.1cm 1cm 0.6cm},clip]{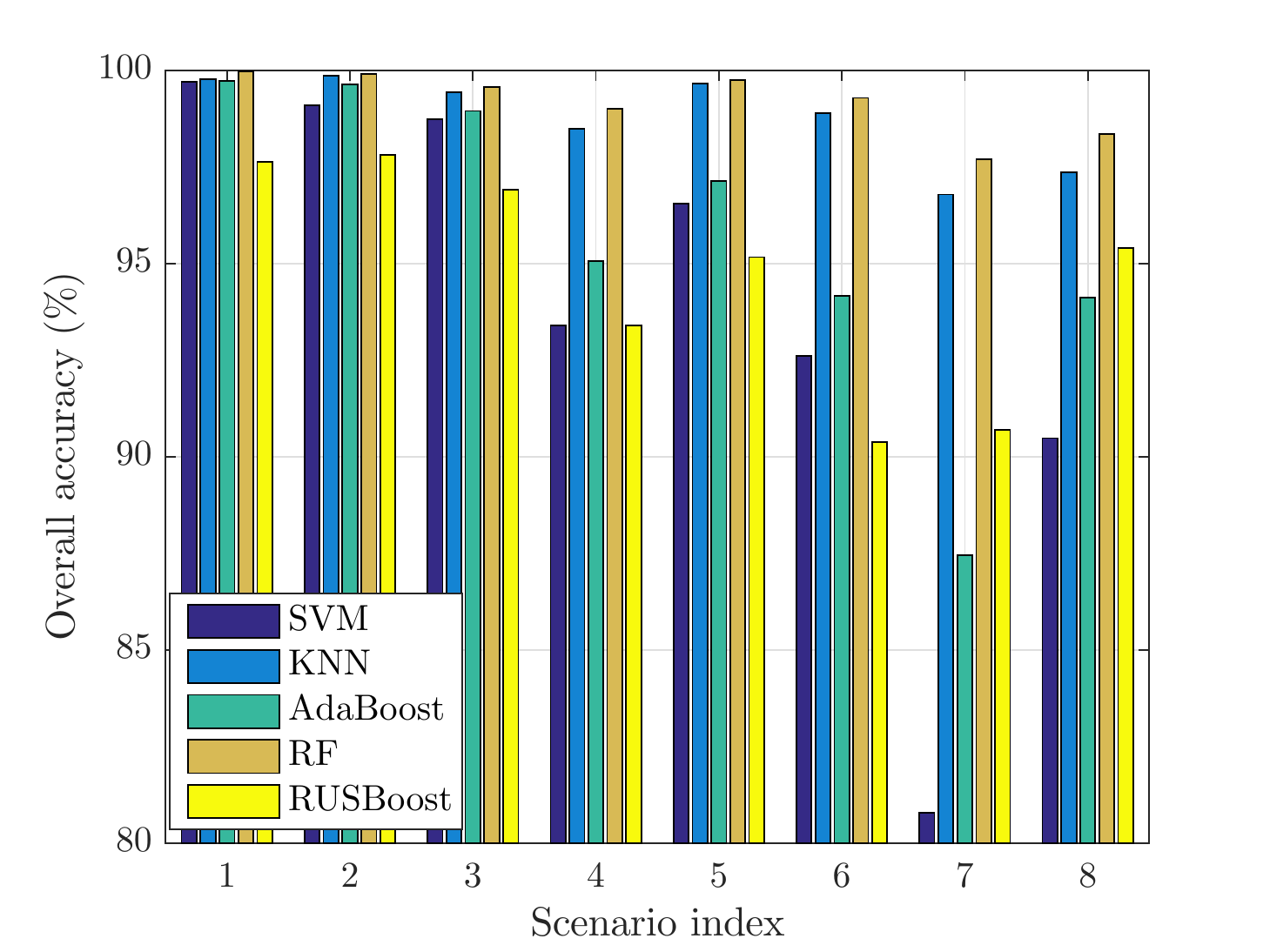} 
	\caption{Overall accuracy per scenario for buffer state classification by 10-fold cross-validation}
	\label{fig:BufferStates_accuracyScenario}  
\end{figure}
	
The respective confusion matrix is presented in Table \ref{table:ConfusionMatrix_Buffer}, where again we highlight the performance of RF as the best algorithm. From this table we can see that SVM and AdaBoost mainly suffer from falsely classifying a depleting state as a steady state.  It is interesting to observe that this is not true for RUSBoost which does not fall below 91\% in terms of correct prediction per class, compared to 83.0\% and 86.5\% for SVM and AdaBoost respectively. The reason is that RUSBoost gives more weight to classes with few samples, which hurts the correct classification of the dominant steady state class (see Table \ref{table:nSamples}) and therefore the overall accuracy, as we verify in \figurename{ \ref{fig:BufferStates_accuracy}}. As explained before, post-processing methods correcting isolated samples that are clearly misclassified can improve the accuracy results for this problem as well. 
	
\figurename{ \ref{fig:BufferStates_accuracyScenario}} shows the overall accuracy of each ML algorithm per experimental scenario. This allows to verify how challenging is buffer state classification for each designed scenario. All algorithms perform better for (s1)-(s2), i.e. the simplest scenarios in our set involving a single filling and steady state. Their performance is slightly reduced for (s3), mainly due to the transition period that follows the manual quality change. (s4)-(s8) are clearly more challenging, since more buffer state changes are introduced. Nevertheless, RF and KNN manage to maintain an accuracy higher than 97.7\% and 96.8\%, respectively, for all scenarios.
	
\begin{figure}[!t]
	\centering
	\includegraphics[width=1\linewidth,trim={0.2cm 0.1cm 1cm 0.6cm},clip]{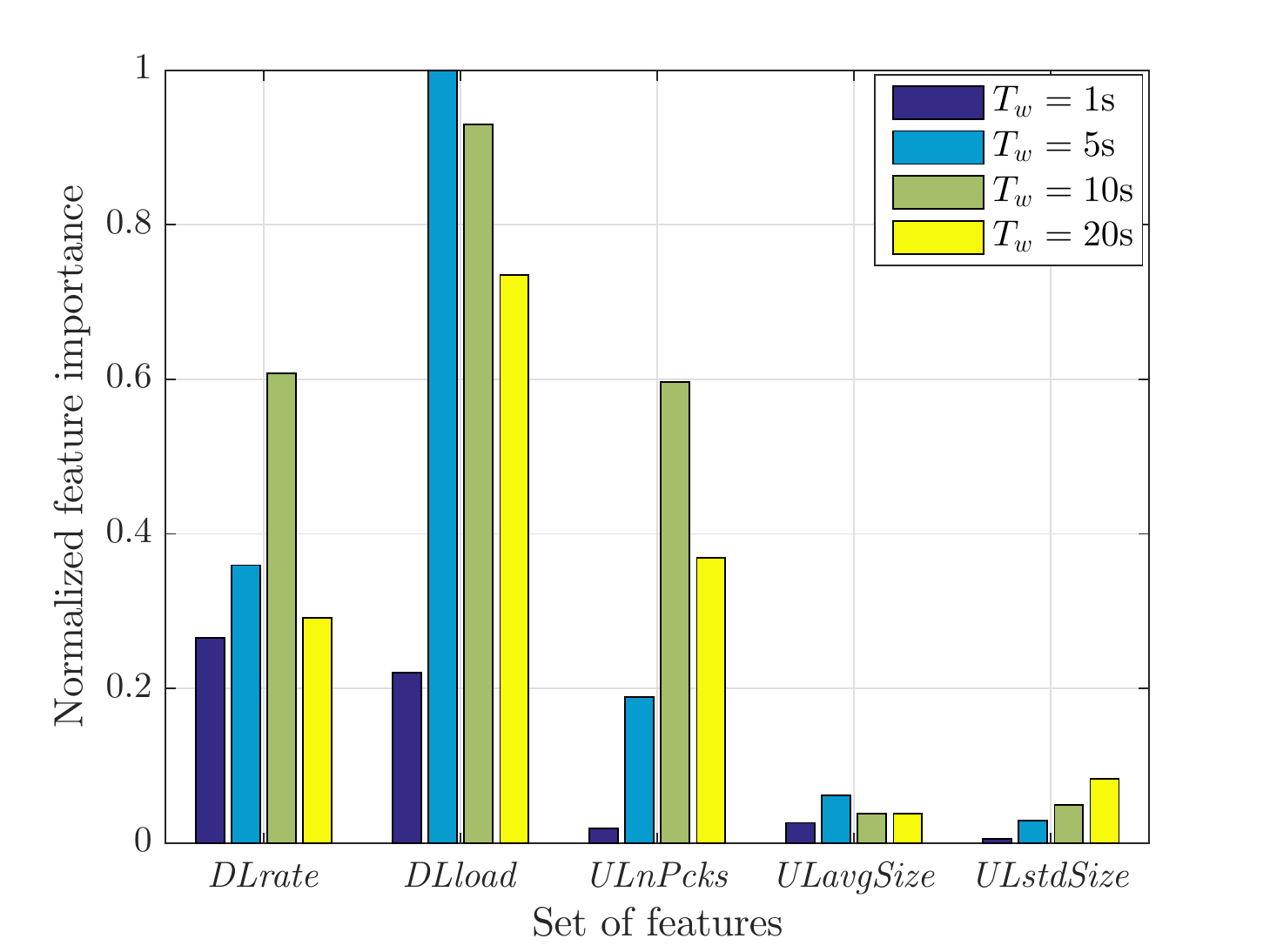} 
	\caption{Feature importance for buffer state classification using RF and 10-fold cross-validation}
	\label{fig:BufferStates_importance}  
\end{figure}		
	
\figurename{ \ref{fig:BufferStates_importance}} presents the normalized importance of each feature using the RF algorithm. 
As before, we apply $L=4$ sliding windows in parallel in order to capture both short-term and long-term variations. Contrary to \figurename{ \ref{fig:FlowID_importance}}, here we can see that \emph{DLload} is the most important feature, with significant contributions from \emph{DLrate} and \emph{ULnPckts}. This comes as no surprise, as these are the 3 features that we specifically selected for HAS traffic. Combined information about downlink streaming rate, percentage of time used for streaming data and frequency of uplink packets enables RF to clearly distinguish different buffer states and achieve an overall accuracy of 99.3\%. \figurename{ \ref{fig:BufferStates_importance}} also justifies our selection of multiple sliding windows, with important features for all $T_w$ values.

\begin{figure}[t]
	\centering
	\includegraphics[width=1\linewidth,trim={0.2cm 0.1cm 1cm 0.6cm},clip]{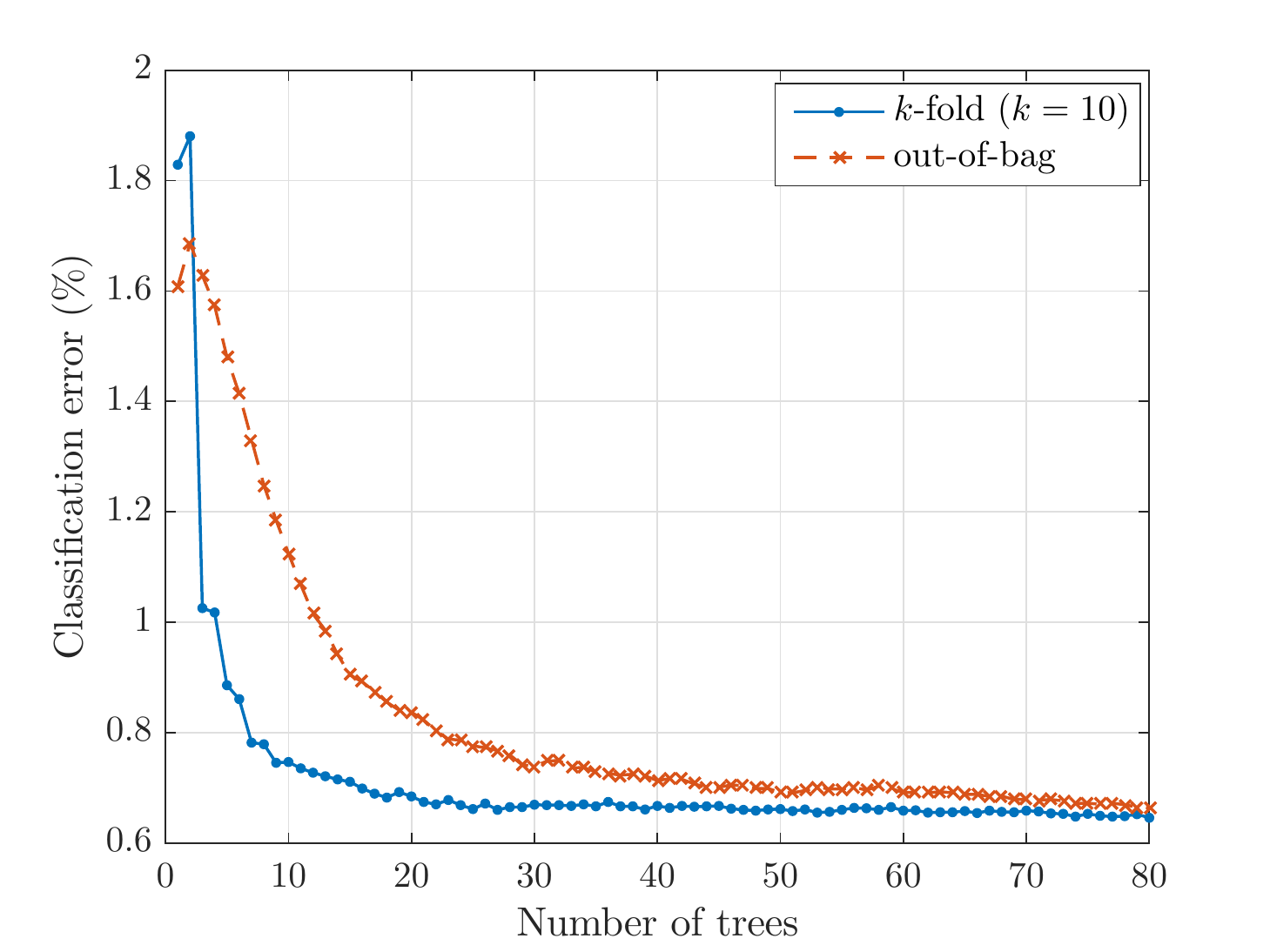} 
	\caption{Classification error for buffer state classification as a function of the number of trees using RF}
	\label{fig:BufferStates_nTrees}  
\end{figure}	
	
\begin{table}[!t]
	\renewcommand{\arraystretch}{1.3}
	\caption{Runtime for buffer state classification (mean $\mu$ and standard deviation $\sigma$)}  
	\label{table:CompTime}
	\centering
	\begin{tabular}{p{0.17\columnwidth}p{0.12\columnwidth}p{0.12\columnwidth}p{0.12\columnwidth}p{0.12\columnwidth}} 
		\hline
		\bfseries ML alg.  & \multicolumn{2}{l}{\bfseries Training} & \multicolumn{2}{l}{\bfseries Prediction}  \\ 
		& $\mu$ (s) & $\sigma$ (s) & $\mu$ (ms) & $\sigma$ (ms) \\ \hline
		SVM 		& 950.99 & 2.37 & 63.73  & 0.31 \\
		KNN 		& 0.96	 & 0.04 & 358.25 & 5.03 \\
		AdaBoost 	& 9.94 	 & 0.10 & 13.25  & 0.68 \\
		RF 			& 21.02  & 0.22 & 19.73  & 1.50 \\
		RUSBoost 	& 6.11   & 0.24 & 13.38  & 1.33 \\	\hline
	\end{tabular}
\end{table}	
	
\figurename{ \ref{fig:BufferStates_nTrees}} presents the
classification error of RF as a function of the number of trees, which is a key parameter for its implementation. Both out-of-bag  and $k$-fold cross-validation error are shown to have similar behavior, quickly decreasing as the number of trees increases. This trend holds up to a value around 50 trees where both errors  converge and stabilize close to a value of 0.7\% and thus, training more trees does not increase performance. We can also verify that even our initial choice of 30 trees has only a slight impact on the out-of-bag error with an increase of 0.03\%.  
	
Finally, Table \ref{table:CompTime} presents the runtime performance of the ML algorithms. These measurements were obtained in the same manner as the runtime for video flow classification (Table \ref{table:CompTimeFlow}) and are similar to these results. We notice that SVM is the most demanding in training time, KNN suffers in prediction time, while RF, AdaBoost and RUSBoost still perform well in both prediction and training phases. We see that the runtime is generally higher than in Table \ref{table:CompTimeFlow}, and even doubles for some cases. This verifies that our buffer state classification problem is computationally more demanding than the presented flow classification, due to the higher number of classes.  

\section{Conclusion}
\label{sec:conclusions}
We introduced a new traffic profiling system to classify flows and buffer states of HAS traffic in real time, based on machine learning. The core of our approach is a classifier that separates HAS from non-HAS traffic and detects 4 buffer states of the streaming client. 

Studying 5 classification methods in our dataset for YouTube's mobile streaming client shows that separating HAS from Web traffic and file downloads is not a challenging problem. Even with highly varying link quality and video bit-rate adaptation, all ML models closely approached 100\% accuracy. For buffer state classification, however, SVM and boosting methods failed, making RF and KNN a very accurate choice in general, with RF being a most attractive approach both in terms of accuracy and runtime. 

Our probably most surprising finding is that such high accuracy can be reached with a small, generic feature set that is only observed at the IP layer. Since no transport layer information is used, our approach works equally for TCP and UDP-based HAS traffic (e.g., with QUIC). Since no application layer information is used, our system does not interfere with end-to-end encryption and requires neither DPI nor cross-layer signaling. Since the feature set is small, on-line and off-line complexity are consistently low. Traffic profiling, thus, provides MNOs with a low-complexity alternative to DPI-based packet dissection and OTT flags.

As buffer states, and their manifestation in packet IATs, are a fundamental property of media streaming, we believe that our feature set will maintain high accuracy for various major streaming services, independently of the content. Our future work will, thus, prioritize the extension of our high quality dataset to more videos and to more services than YouTube. We plan to publish our data set in the near future, so that other researchers can reproduce our results and use our measurements in their own studies. 
  
\section*{Acknowledgments}
We thank Florian Wamser, Bernd Zeidler, Michael Seufert, and Phuoc Tran-Gia from the University of W{\"u}rzburg, Germany for their valuable contribution on the development of the YouTube wrapper application and for their comments that greatly improved the design of our experiments. We also thank our colleagues Yuejun Wei and Yuchen Chen for their insight on the product integration of our work.

\bibliography{IEEEabrv,TP_ML-arxiv}
\bibliographystyle{IEEEtran}

\end{document}